\documentclass[aps,prl,10pt,twocolumn,superscriptaddress]{revtex4-1}

\usepackage{CJK}
\usepackage{graphicx}
\usepackage[verbose=true]{epstopdf}
\usepackage{physics} 
\usepackage{hyphenat} 
\usepackage[bookmarks=false,bookmarksopen=false,pdfpagelayout=SinglePage,pdfstartview=Fit]{hyperref}
\hypersetup{colorlinks,breaklinks,linkcolor=blue,menucolor=green,urlcolor=blue,citecolor=blue}
\usepackage[dvipsnames]{xcolor}

\begin{document}
\begin{CJK*}{GB}{}

\title{Singlet Pathway to the Ground State of Ultracold Polar Molecules}

\author{A.~Yang}
\affiliation{Centre for Quantum Technologies (CQT), 3 Science Drive 2, Singapore 117543}
\author{S.~Botsi}
\affiliation{Centre for Quantum Technologies (CQT), 3 Science Drive 2, Singapore 117543}
\author{S.~Kumar}
\affiliation{Centre for Quantum Technologies (CQT), 3 Science Drive 2, Singapore 117543}
\author{S.~B.~Pal}
\affiliation{Centre for Quantum Technologies (CQT), 3 Science Drive 2, Singapore 117543}
\author{M.~M.~Lam}
\affiliation{Centre for Quantum Technologies (CQT), 3 Science Drive 2, Singapore 117543}
\author{I.~\v{C}epait\.{e}}
\affiliation{Centre for Quantum Technologies (CQT), 3 Science Drive 2, Singapore 117543}
\author{A.~Laugharn}
\affiliation{Centre for Quantum Technologies (CQT), 3 Science Drive 2, Singapore 117543}
\author{K.~Dieckmann}
\email[Electronic address:]{phydk@nus.edu.sg}
\affiliation{Centre for Quantum Technologies (CQT), 3 Science Drive 2, Singapore 117543}
\affiliation{Department of Physics, National University of Singapore, 2 Science Drive 3, Singapore 117542}

\date{\today}

\begin{abstract}
Starting from weakly bound Feshbach molecules, we demonstrate a two\hyp{}photon pathway to the dipolar ground state of bi\hyp{}alkali molecules that involves only singlet\hyp{}to\hyp{}singlet optical transitions. This pathway eliminates the search for a suitable intermediate state with sufficient singlet\hyp{}triplet mixing and the exploration of its hyperfine structure, as is typical for pathways starting from triplet dominated Feshbach molecules. By selecting a Feshbach state with a stretched singlet hyperfine component and controlling the polarization of the excitation laser, we assure coupling to only a single hyperfine component of the $\textrm{A}^{1}\Sigma^{+}$ excited potential, even if the hyperfine structure is not resolved. Similarly, we address a stretched hyperfine component of the $\textrm{X}^{1}\Sigma^{+}$ rovibrational ground state, and therefore an ideal three level system is established. We demonstrate this pathway with ${}^{6}\textrm{Li}{}^{40}\textrm{K}$ molecules. By exploring deeply bound states of the $\textrm{A}^{1}\Sigma^{+}$ potential, we are able to obtain large and balanced Rabi frequencies for both transitions. This method can be applied to other molecular species.
\end{abstract}

\maketitle
\end{CJK*}
The rich physics of ultracold dipolar quantum gases \cite{Trefzger2011,Baranov2012} based on the long\hyp{}range and anisotropic interaction has moved into the focus of many experimental research activities. Studies on Rydberg atoms \cite{Schauss2015,Labuhn2016} and recent results with strongly magnetic atoms \cite{Boettcher2019,Chomaz2019} represent two important realizations of dipolar systems. A third approach is using polar molecules, which combine strong dipolar interactions and long lifetimes when trapped in deep optical lattices or low dimensional traps. Therefore, they are prominent candidates for research in the areas of quantum simulation \cite{Micheli2006,Gadway2016} and quantum information \cite{deMille2002,Yelin2006,Andre2006aca}. Moreover, polar molecules are an important platform for metrology \cite{Kozlov1995,Hudson2006,Andreev2018} and are a testbed for ultracold chemistry \cite{Carr2009,Bell2009,Quemener2012,Jin2012}. For the difficult task of producing trapped ultracold molecular samples several methods were explored \cite{Wynar2000,Bethlem2000,Kraft2006,Shuman2010}, and over the recent years much progress was made with laser cooling of molecules from a buffer gas source \cite{Anderegg2018}. However, the highest numbers and densities and lowest temperatures for ultracold molecular samples were produced by association of ultracold atoms via magnetically tunable Feshbach resonances \cite{Koehler2006}. To access large dipole moments, such weakly bound molecules need to be transferred into their rovibrational ground state by stimulated Raman adiabatic passage (STIRAP) \cite{Vitanov2017}. The ground state was achieved for several heteronuclear bi\hyp{}alkali combinations \cite{Ni2008,Takekoshi2014,Molony2014,Park2015,Guo2016}. This bottom\hyp{}up approach is the only one that recently reached the remarkable milestone of quantum degeneracy for the case of KRb \cite{DeMarco2019}. To obtain molecules with higher ground state dipole moments other bi\hyp{}alkali combinations with larger mass ratios are being investigated. Obtaining efficient ground state transfer requires as a cornerstone studying the molecular structure and identifying a suitable electronically excited state. Most experiments so far follow the strategy that was pioneered in KRb \cite{Ni2008} by using an intermediate excited state of mixed singlet and triplet character. This facilitates a coherent transfer from a Feshbach molecule state with triplet character to the singlet ground state. To gain full control over the molecular state, spectroscopically resolved hyperfine components of the excited state need to be identified, as coupling to unresolved hyperfine levels will lead to a low STIRAP efficiency \cite{Vitanov1999}. Identifying such states typically requires extensive spectroscopic surveys \cite{Kotochigova2009,Debatin2011,Zabawa2011,Park2015b,Guo2017,Rvachov2018} to find states where the triplet admixture leads to large hyperfine splitting, while the singlet admixture is strong enough to address the ground state. Moreover, even if the hyperfine structure is resolved, off\hyp{}resonant coupling to other hyperfine components of the intermediate state can lead to an undesired superposition of several hyperfine components of the ground state \cite{Liu2019}.

\begin{figure}[t]
	\includegraphics[width=8.6cm]{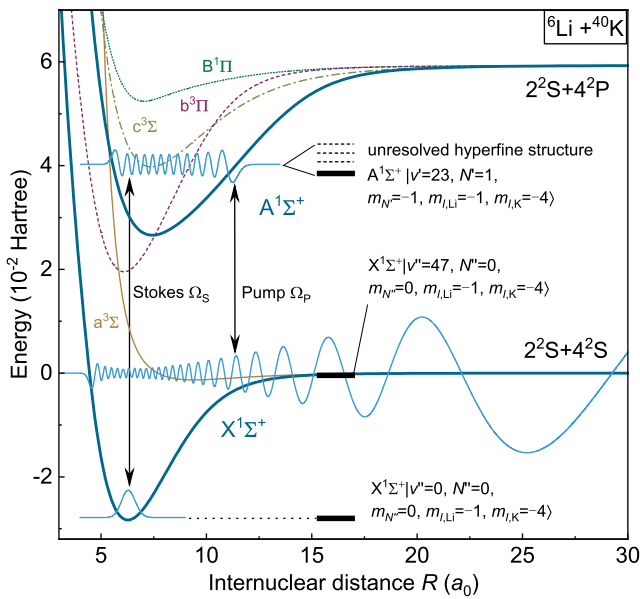}
	\caption{\label{fig:PECs} Potential energy curves of ${}^{6}\textrm{Li}{}^{40}\textrm{K}$ molecules. The pump beam is used to couple the singlet component of the Feshbach molecules ($\textrm{X}^{1}\Sigma^{+}\ket{v''=47}$) to a deeply bound vibrational state of the $\textrm{A}^{1}\Sigma^{+}$ potential with Rabi frequency $\Omega_{\mathrm{p}}$. From there the Stokes beam couples to the rovibrational ground state in the $\textrm{X}^{1}\Sigma^{+}$ potential with $\Omega_{\mathrm{S}}$.}
\end{figure}

In this letter, we present a pathway to the ground state that avoids the use of perturbed potentials by only using singlet\hyp{}to\hyp{}singlet transitions for both pump and Stokes couplings as illustrated in Fig.~\ref{fig:PECs}. For addressing the $\textrm{X}^{1}\Sigma^{+}$ rovibrational ground state of the molecules, the $\textrm{A}^{1}\Sigma^{+}$ potential is explored to search for a suitable intermediate state. Here, we demonstrate that by making use of deeply bound vibrational states in this potential, we are able to achieve sufficient Franck\hyp{}Condon factors (FCF) to the ground state with moderate laser powers. The use of deeply bound vibrational states has the advantage that states of mixed singlet\hyp{}triplet character are less frequent, due to the larger level spacing. To allow for the excitation to the unperturbed $\textrm{A}^{1}\Sigma^{+}$ potential, it is necessary to employ a Feshbach resonance with significant admixture from the singlet ground state potential. For the $\textrm{A}^{1}\Sigma^{+}$ potential, its hyperfine structure is typically not resolved due to the absence of spin and internal magnetic field. We then demonstrate how to address a sole hyperfine component of the ground state, even if the hyperfine structure of the excited state is not resolvable. Our pathway is facilitated by selecting the Feshbach resonance, such that the singlet admixture to the Feshbach molecular state consists of only one hyperfine component. This singlet component corresponds to a fully stretched state of the nuclear spin projections. Starting from the stretched state and applying circularly polarized spectroscopy light ensures that only one ground state hyperfine component can be addressed. In the following, this scheme is applied to ${}^{6}\textrm{Li}{}^{40}\textrm{K}$ molecules. We describe the selection of a suitable Feshbach state and present our spectroscopic results for several deeply bound vibrational states of the $\textrm{A}^{1}\Sigma^{+}$ potential as well as our measurements of the ground state by two\hyp{}photon spectroscopy. 

Our experiments start from a quantum degenerate mixture of $10^{5}$ ${}^{6}\textrm{Li}$ and $8\times 10^{4}$ ${}^{40}\textrm{K}$ atoms in a magnetic trap, which is sympathetically cooled via evaporative cooling of ${}^{87}\textrm{Rb}$ \cite{Taglieber2008}. After expunging ${}^{87}\textrm{Rb}$ from the magnetic trap, the Fermi\hyp{}Fermi mixture is transferred into a crossed optical dipole trap. This is followed by preparation of suitable atomic hyperfine states and magneto\hyp{}association of typically $10^{4}$ Feshbach molecules \cite{Voigt2009}. The molecules are detected by absorption imaging in the presence of the magnetic field after Feshbach molecular association, where atoms and weakly bound molecules can be simultaneously detected. We reduce the background due to remaining unbound ${}^{6}\textrm{Li}$ atoms by transferring them to a different Zeeman level, which is off\hyp{}resonant to the imaging light.

For the molecular association in previous work we employed a Feshbach resonance at $15.54\,$mT \cite{Voigt2009}. To select an initial molecular state for the spectroscopy that contains a sole singlet admixture, we use a different Feshbach resonance at $21.56\,$mT. This is achieved by preparing the lithium and potassium atoms in the $\ket{F_{\mathrm{Li}}=\frac{1}{2},m_{F,\mathrm{Li}}=-\frac{1}{2}}$ and $\ket{F_{\mathrm{K}}=\frac{9}{2},m_{F,\mathrm{K}}=-\frac{9}{2}}$ hyperfine states respectively, where $F$ is the hyperfine quantum number, and $m_{F}$ is the projection of hyperfine angular momentum on the quantization axis. The sum of the projection quantum numbers is $M=-5$, which is conserved during molecule formation. As most Feshbach resonances of the ${}^{6}\textrm{Li}-{}^{40}\textrm{K}$ mixture are narrow \cite{Wille2008}, the closed molecular channel should be considered. In the molecular basis $\ket{S,m_S,m_{I,\mathrm{Li}},m_{I,\mathrm{K}}}$, the total projection quantum number then equals to $M = m_{S} + m_{I,\mathrm{Li}}+m_{I,\mathrm{K}} = -5$. Hence, only one spin singlet and three spin triplet states contribute to the closed channel: $\ket{0,0,-1,-4}$, $\ket{1,0,-1,-4}$, $\ket{1,-1,0,-4}$, and $\ket{1,-1,-1,-3}$. Since the nuclear spin quantum numbers are $I_{\mathrm{Li}}=1$ and $I_{\mathrm{K}}=4$, the only contributing spin singlet state is fully stretched. The singlet and triplet states are mixed by the hyperfine coupling, which can be well described by the Asymptotic Bound State Model \cite{Tiecke2010}. Based on this, we estimate the admixture of the spin singlet state to the molecular eigenstate at the Feshbach resonance to be $52$\% \cite{supplement}. This significant singlet character of the Feshbach state is an excellent starting point for the spectroscopy of $\textrm{A}^{1}\Sigma^{+}$.

To search for a suitable intermediate state for locating the ground state, we study the excitation from the initial Feshbach state $\textrm{X}^{1}\Sigma^{+}\ket{v''=47,J''=0}$ to the rotational excited states in $\textrm{A}^{1}\Sigma^{+}$ potential $\textrm{A}^{1}\Sigma^{+}\ket{v',J'=1}$ as shown in Fig.~\ref{fig:PECs}, where $v$ is the vibrational quantum number. Since only $^{1}\Sigma$ states are involved and the $J''=0\rightarrow J'=0$ transition is forbidden for the total angular momentum $J$ of the molecules, the quantum number $N$ for molecular rotation is not conserved. Hence, starting from a Feshbach state with $N''=0$, we can only probe excited states of $\textrm{A}^{1}\Sigma^{+}$ with $N'=1$. As a consequence, the rotational angular momentum projection $m_N$ needs to be considered. In the extended uncoupled molecular basis $\ket{S,m_S,m_{I,\mathrm{Li}},m_{I,\mathrm{K}}, N, m_N}$ the following three states are possible $\ket{0,0,-1,-4,1,0}$, $\ket{0,0,0,-4,1,-1}$, $\ket{0,0,-1,-3,1,-1}$, if we drive $\pi$ transitions and the total projection $M= m_{I,\mathrm{K}}+m_{I,\mathrm{Li}}+m_N = -5$ is conserved.  These states can be mixed due to hyperfine coupling originating from nuclear\hyp{}spin rotation, nuclear quadrupole, and the nuclear\hyp{}spin dipole interactions. In the $\textrm{A}^{1}\Sigma^{+}$ potential the hyperfine coupling constants are expected to be small \cite{supplement} and the hyperfine structure is not resolved. Therefore, the excitation can couple to different hyperfine components. To achieve excitation to the sole singlet hyperfine component $\ket{0,0,-1,-4,1,-1}$ in the $\textrm{A}^{1}\Sigma^{+}$ potential, we apply $\sigma^{-}$\hyp{}polarized light. As stretched states do not couple, this argument is also valid, if the hyperfine coupled basis is considered. Since our Feshbach state is in $\ket{F''=5,m_{F}''=-5}$, the addressable excited hyperfine states are $F'=4,5,6$. With $\sigma^{-}$ polarization we can address only the $\ket{F'=6,m_{F}'=-6}$ stretched state.

\begin{figure}[t]
	\includegraphics[width=8.6cm]{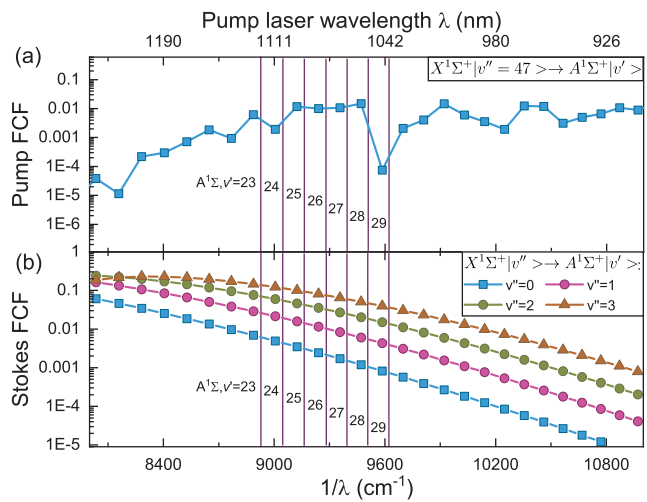}
	\caption{\label{fig:FCF} Franck\hyp{}Condon overlap factors (FCF) of the vibrational states of the $\textrm{A}^{1}\Sigma^{+}$ potential with (a) the molecular Feshbach state and (b) the lowest vibrational states of the $\textrm{X}^{1}\Sigma^{+}$ potential. Detail of the calculation of the FCFs is given in the supplemental material \cite{supplement}. The vertical lines indicate the wavelengths at which we identified vibrational states in the experiment. The lines connecting the points are a guide to the eye. }
\end{figure}
A suitable intermediate vibrational state of the $\textrm{A}^{1}\Sigma^{+}$ potential needs to have a good FCF with the Feshbach molecular state as well as the vibrational ground state of the $\textrm{X}^{1}\Sigma^{+}$ potential. On the one hand, the overlap with the ground state is favorable, if deeply bound vibrational states at long wavelengths are used, as shown in Fig.~\ref{fig:FCF}(b). On the other hand, at long wavelengths the good overlap with the singlet component of the closed channel dominated Feshbach molecular state ceases, as inferred from Fig.~\ref{fig:FCF}(a). In the experiment we cover a wavelength range of $1038\,$nm - $1120\,$nm, where good FCFs for both Pump and Stokes transitions are expected. It should be noted that the variation of the transition strength throughout this range is determined by the vibrational wave\hyp{}function overlap. This is because for the $\textrm{X}^{1}\Sigma^{+}\rightarrow\textrm{A}^{1}\Sigma^{+}$ transition the transition dipole matrix element (TDM) does not vary significantly with the internuclear distance, as we find from our ab\hyp{}initio calculation \cite{supplement}.

\begin{figure}[t]
\includegraphics[width=8.6cm]{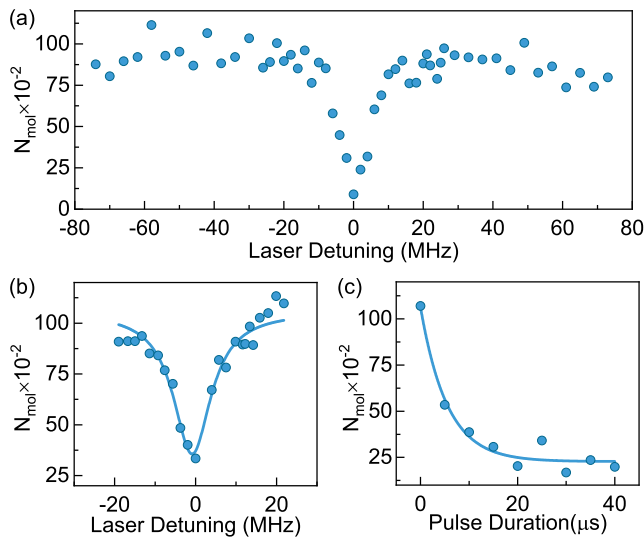}
\caption{\label{fig:onephoton} Spectroscopy of the $\textrm{A}^{1}\Sigma^{+}\ket{v'=23}$ excited state. The transition resonance is reflected as molecule number losses. (a) Broad spectrum to exclude hyperfine structure originated from accidental perturbation caused by $\textrm{b}^{3}\Pi$; (b) Spectrum with reduced laser power; (c) The decay data for resonant irradiation.}
\end{figure}

We perform one\hyp{}photon spectroscopy for the $\textrm{A}^{1}\Sigma^{+}$ vibrational levels by switching off the optical dipole trap and applying laser pulses during time\hyp{}of\hyp{}flight (TOF) before imaging. We use a gain-chip laser diode, which is stabilized by a tunable frequency offset lock to a high\hyp{}finesse optical resonator. The laser frequency is determined with $1\,$MHz resolution using a beat setup with an optical frequency comb system. Our identification of the transition lines is facilitated by previous polarization labeling spectroscopy in a heat pipe \cite{Grochola2012}. By mass scaling of the Dunham coefficients determined by this work to the ${}^{6}\textrm{Li}{}^{40}\textrm{K}$ isotopologue \cite{supplement}, we are able to address individual vibrational states. Here, the large level separation occurring for deeply bound states leads to the unambiguous assignment of the vibrational level index. The locations of seven deeply bound excited states are measured with highly resolving spectroscopy, and given in Tab.~\ref{tab:lines}. As an example, the spectrum for the transition to $\textrm{A}^{1}\Sigma^{+}\ket{v'=23}$ is shown in Fig.~\ref{fig:onephoton}(a). The laser intensity is reduced to avoid line broadening. From simultaneous curve fitting to the observed spectrum, Fig.~\ref{fig:onephoton}(b), and the exponential decay of the molecular number with the irradiation time, Fig.~\ref{fig:onephoton}(c), the transition strength is inferred from the normalized Rabi frequency $\bar{\Omega}_{\mathrm{p}}$. Consistent with the prediction of the FCFs shown in Fig.~\ref{fig:FCF}(a), we do not observe a reduction of the transition strength of this state as compared with the strengths measured at lower wavelengths as summarized in Tab.~\ref{tab:lines}. 

\begin{table}[b]
\centering
	\begin{ruledtabular}
			\begin{tabular}{ccc}
			\multicolumn{3}{l}{\hspace{2.3cm}$\textrm{X}^{1}\Sigma^{+}\ket{v''=47}\rightarrow\textrm{A}^{1}\Sigma^{+}\ket{v'}$}\\[3pt]
			$v'$ & $f$ (THz) & $\bar{\Omega}_{\mathrm{P}}$ (kHz/$\mathrm{\sqrt{mW/cm^{2}}}$)\\[3pt]
			\hline
			23 & 267.842232(1) & 2$\pi\times 2.3(2)$\\
			24 & 271.383075(1) & 2$\pi\times 1.6(2)$\\
			25 & 274.898734(2) & 2$\pi\times 5.1(2)$\\
			26 & 278.388059(2) & 2$\pi\times 0.7(1)$\\
			27 & 281.846301(1) & 2$\pi\times 8.8(3)$\\
			28 & 285.281804(1) & 2$\pi\times 3.0(3)$\\
			29 & 288.688289(1) & 2$\pi\times 3.1(3)$\\
		\end{tabular}{}
		\end{ruledtabular}
	\caption{\label{tab:lines}Measured resonant frequencies and normalized Rabi frequencies of transitions from $\textrm{X}^{1}\Sigma^{+}\ket{v''=47}$ to deeply bound $\textrm{A}^{1}\Sigma^{+}\ket{v'=23-29}$ excited states.}
	\end{table}

An important feature of all observed lines is the absence of hyperfine\hyp{}structure for the measured linewidth of $5\,$MHz, if the excitation is driven by $\sigma^-$ light. This is expected as the stretched hyperfine state is the only available final state. An observable hyperfine structure is expected to occur for spin\hyp{}triplet states. In this case, the atomic hyperfine splitting for potassium atoms of a few tens of MHz can be used as an estimate for the molecular case. Since no hyperfine structure is observed over the $160\,$MHz span of the measurement shown in Fig.~\ref{fig:onephoton}(a), it can be excluded that the observed vibrational states are perturbed by a significant contribution of the $\textrm{b}^{3}\Pi$ states via spin\hyp{}orbit coupling \cite{supplement}. Further, in similar measurements no hyperfine structure is resolved, if the excitation is driven with $\pi$\hyp{}polarized light, which allows to address several hyperfine states. This is consistent with estimates of the dominant quadrupole hyperfine interaction for the $\textrm{A}^{1}\Sigma^{+}$ potential based on ab\hyp{}initio calculations \cite{supplement}. 

To address the low lying vibrational states of the $\textrm{X}^{1}\Sigma^{+}$ ground state, we make use of the stretched hyperfine state $\ket{0,0,-1,-4,1,-1}$ of the $\textrm{A}^{1}\Sigma^{+}$ potential as an intermediate state with projection quantum number $M' = -6$. For the rovibrational ground state with $N'' = 0$, only the stretched hyperfine component with $M''=-5$ can couple to the excited state. In this case $\sigma^{-}$\hyp{}polarized Stokes light needs to be applied. Consequently, off\hyp{}resonant coupling to other hyperfine states is prevented for both transitions.

We perform two\hyp{}photon spectroscopy during TOF to determine the energy difference between the rovibrational ground state and the intermediate state. For the spectroscopy, a pump pulse duration of $18\,\mathrm{\mu s}$ is used and the Stokes pulse is switched on and off $2\,\mathrm{\mu s}$ earlier and respectively later than the pump pulse. 

The search for the ground state levels is assisted by the $\textrm{X}^{1}\Sigma^{+}$ potential curve of \cite{Tiemann2009}. We first identify the transition frequencies to $\textrm{X}^{1}\Sigma^{+}\ket{v''=3}$ by scanning the Stokes laser detuning, while keeping the pump laser on resonance with respect to the transition to $\textrm{A}^{1}\Sigma^{+}\ket{v'=29}$. The Stokes light is derived from a low power diode laser system. As shown in Fig.~\ref{fig:FCF}(b), lower vibrational states have significantly smaller FCFs. Therefore, it is impossible to address the $\textrm{X}^{1}\Sigma^{+}\ket{v''=0}$ with the same intermediate state, unless laser power of several Watts is available. 

To facilitate the search via other intermediate vibrational levels we make use of a tunable dye laser system (Coherent 699). The DCM dye in use covers a wavelength range of $640\hyp{}690\,$nm with a power up to $300\,$mW applied to the molecules. The frequency of the dye laser is controlled by our previously described interferometric laser frequency stabilization \cite{Brachmann2012}, which allows continuous in\hyp{}lock scanning of the frequency over many gigahertz in steps of $1\,$MHz. A more deeply bound intermediate state $\textrm{A}^{1}\Sigma^{+}\ket{v'=23}$ is utilized to access the low lying vibrational states including the $\textrm{X}^{1}\Sigma^{+}\ket{v''=0}$ with sufficient FCFs. All measured transition frequencies are summarized in Tab.~\ref{tab:lines2}(b). The predicted frequencies deviate from the measured values by approximately $3\,$GHz, which is within the uncertainty of $6\,$GHz given in \cite{Tiemann2009}. 

\begin{table}[t]
	\centering
	\begin{ruledtabular}
		\begin{tabular}{ccc}
			\multicolumn{3}{l}{(a)\hspace{2.3cm}$\textrm{A}^{1}\Sigma^{+}\ket{v'=29}\rightarrow\textrm{X}^{1}\Sigma^{+}\ket{v''}$}\\[3pt]
			$v''$ & $f$ (THz) & $\bar{\Omega}_{\mathrm{S}}$ (kHz/$\mathrm{\sqrt{mW/cm^{2}}}$)\\[3pt]
			\hline
			2 & 458.389566(5) & 2$\pi\times 43(8) $\\
			3 & 451.871767(10) & 2$\pi\times 152(52) $\\
			\hline\\
			\multicolumn{3}{l}{(b)\hspace{2.3cm}$\textrm{A}^{1}\Sigma^{+}\ket{v'=23}\rightarrow\textrm{X}^{1}\Sigma^{+}\ket{v''}$}\\[3pt]
			$v''$ & $f$ (THz) & $\bar{\Omega}_{\mathrm{S}}$ (kHz/$\mathrm{\sqrt{mW/cm^{2}}}$)\\[3pt]
			\hline
			0 & 450.841975(2) & 2$\pi\times 12(2) $\\
			1 & 444.148838(5) & 2$\pi\times 84(18) $\\
			2 & 437.543516(8) & 2$\pi\times 256(59) $\\
		\end{tabular}{}
	\end{ruledtabular}
	\caption{\label{tab:lines2}Measured resonance frequencies and normalized Rabi freqencies for the coupling (a) between $\textrm{A}^{1}\Sigma^{+}\ket{v'=29}$ and low lying $\textrm{X}^{1}\Sigma^{+}\ket{v''=2-3}$; (b) between $\textrm{A}^{1}\Sigma^{+}\ket{v'=23}$ and $\textrm{X}^{1}\Sigma^{+}\ket{v''=0-2}$ using $\pi$ polarization as initially the hyperfine coupling of excited states is not considered.}
\end{table}

\begin{figure}[t]
	\includegraphics[width=8.6cm]{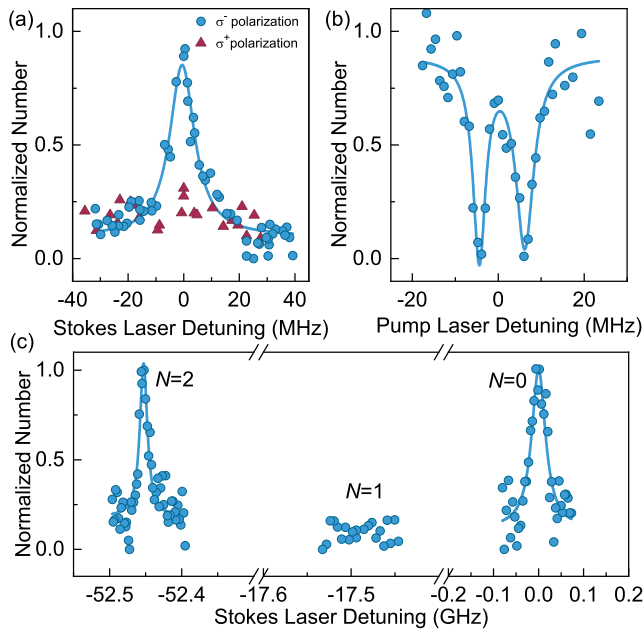}
	\caption{\label{fig:twophoton}  Spectroscopy of $\textrm{X}^{1}\Sigma^{+}\ket{v''=0}$ ground state using $\textrm{A}^{1}\Sigma^{+}\ket{v'=23, N'=1}$ as an intermediate state: (a) The pump laser with $\sigma^{-}$ polarization is tuned to one photon resonance. The Stokes\hyp{}laser frequency is scanned using $\sigma^{-}$ and $\sigma^{+}$ polarization. (b) Both pump and Stokes beams are $\sigma^{-}$\hyp{}polarized. The Stokes laser is fixed on two\hyp{}photon resonance, while the pump\hyp{}laser detuning is scanned. (c) Two photon spectroscopy including the rotational excited state $\textrm{X}^{1}\Sigma^{+}\ket{v''=0,N''=2}$: the rotation energy, which is indicated by the energy difference between the rotational ground state and excited state, is given by $E_{N''} = \mathrm{B_{0}}N''(N''+1)$. The transition to $N''=1$ is not observed due to the selection rule $\Delta N=\pm 1$.}
\end{figure}

The results of two\hyp{}photon spectroscopy for measuring the ground state energy for the $\textrm{X}^{1}\Sigma^{+}\ket{v''=0,N''=0}$ state are shown in Fig.~\ref{fig:twophoton}. In Fig.~\ref{fig:twophoton}(a), the Feshbach molecular signal is restored due to the Autler\hyp{}Townes effect, if the Stokes laser is tuned to resonance. If the polarization of the Stokes laser is changed to $\sigma^{+}$, no coupling is expected and therefore no line is observed. Fig.~\ref{fig:twophoton}(b) shows the Autler\hyp{}Townes splitting (ATS) by keeping the Stokes laser on resonance and scanning the pump laser frequency. To confirm the correct assignment of the ground state, we address the rotationally exited $\ket{v''=0,N''=2}$ state, as shown in Fig.~\ref{fig:twophoton}(c). From the separation of the two resonance signals a rotational constant of $\mathrm{B_{0}} = h \times 8.742(3)\,$GHz is inferred.

The transition strengths $\bar{\Omega}_{\mathrm{S}}$ to the $\textrm{X}^{1}\Sigma^{+}$ vibrational states are obtained from the measured widths of the spectra as shown in Fig.~\ref{fig:twophoton}(a) and the applied laser intensities.  The results are shown in Tab.~\ref{tab:lines2}. The scaling of the transition strengths between different lines is in good agreement with the ratio of the respective FCFs shown in Fig.~\ref{fig:FCF}(b). The strong coupling to the $v''=0$ ground state can be inferred from the observed ATS, which shows a Stokes Rabi frequency $\Omega_{\text{S}}$ around $2\pi\times8\,$MHz. $\Omega_{\text{S}}$ of $500\,$kHz can be achieved, even if we switch to a low power diode laser. As the measured Rabi frequency for the pump excitation is of the same order, we have established favorable conditions for a fast STIRAP transfer. 

While our pathway method was demonstrated here for the case of ${}^{6}\textrm{Li}{}^{40}\textrm{K}$, it can be applied to other bi\hyp{}alkali molecules, where a Feshbach state can be found that contains singlet admixture, like LiCs or KCs. 
Other molecular species that are currently studied, like LiYb, RbSr, and CsYb, have only spin\hyp{}doublet ground and excited electronic states. We believe that the method of using a single stretched hyperfine state can be extended to such molecules.

To conclude, we demonstrated a pathway to access the rovibrational ground state of ultracold bi\hyp{}alkali molecules using only singlet states. We found that the $\textrm{A}^{1}\Sigma^{+}$ potential offers good transition strengths to the Feshbach and ground states with moderate laser powers. Starting from a Feshbach state with a stretched hyperfine component of the singlet admixture allows addressing a pure hyperfine component of the ground state, even if the hyperfine structure is not resolved. An ideal three\hyp{}level system is established and is robust against off\hyp{}resonant coupling to other hyperfine components. This pathway has the advantage that it does not require a singlet\hyp{}triplet mixing with fully resolved hyperfine structure in the excited state manifold. Our method is applicable to other types of molecules and represents a simplified and efficient pathway to their ground state.

\begin{acknowledgments}
We thank J.T.M. Walraven for helpful discussions on this manuscript. This research is supported by the National Research Foundation, Prime Ministers Office, Singapore and the Ministry of Education, Singapore under the Research Centres of Excellence program. We further acknowledge funding by the Singapore Ministry of Education Academic Research Fund Tier 2 (grant MOE2015-T2-1-098).
\end{acknowledgments}

%



\end{document}


\begin{CJK*}{}{}

\title{Supplemental Material: Singlet Pathway to the Ground State of Ultracold Polar Molecules}

\author{A.~Yang}
\affiliation{Centre for Quantum Technologies (CQT), 3 Science Drive 2, Singapore 117543}
\author{S.~Botsi}
\affiliation{Centre for Quantum Technologies (CQT), 3 Science Drive 2, Singapore 117543}
\author{S.~Kumar}
\affiliation{Centre for Quantum Technologies (CQT), 3 Science Drive 2, Singapore 117543}
\author{S.~B.~Pal}
\affiliation{Centre for Quantum Technologies (CQT), 3 Science Drive 2, Singapore 117543}
\author{M.~M.~Lam}
\affiliation{Centre for Quantum Technologies (CQT), 3 Science Drive 2, Singapore 117543}
\author{I.~\v{C}epait\.{e}}
\affiliation{Centre for Quantum Technologies (CQT), 3 Science Drive 2, Singapore 117543}
\author{A.~Laugharn}
\affiliation{Centre for Quantum Technologies (CQT), 3 Science Drive 2, Singapore 117543}
\author{K.~Dieckmann}
\email[Electronic address:]{phydk@nus.edu.sg}
\affiliation{Centre for Quantum Technologies (CQT), 3 Science Drive 2, Singapore 117543}
\affiliation{Department of Physics, National University of Singapore, 2 Science Drive 3, Singapore 117542}


\maketitle
\end{CJK*}

\section{Supplemental Material}
In this document we describe methods used for modeling ${}^{6}\textrm{Li}{}^{40}\textrm{K}$ molecules. By means of the asymptotic bound state model we investigate the closed channel state composition of our Feshbach molecules. Then, mass scaling of literature data that allows us to identify spectroscopic transition lines of the $\textrm{A}^{1}\Sigma^{+}$ potential is described. Further, from ab\hyp{}initio calculations we infer the Franck\hyp{}Condon overlap factors, the spin\hyp{}orbit coupling constants, and the relevant hyperfine coupling constants for the interpretation of our spectroscopic measurements.  
\subsection{Asymptotic\hyp{}Bound\hyp{}State Model (ABM)}
\begin{figure}[b]
	\includegraphics[width=8.6cm]{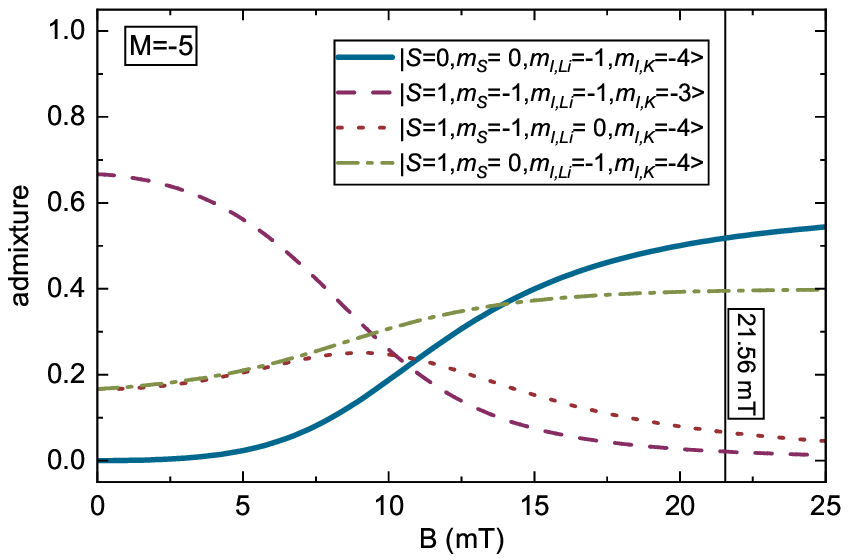}
	\caption{\label{fig:ABM-5} Composition of the resonant molecular eigenstate in terms of singlet and triplet components.}
\end{figure}
The asymptotic\hyp{}bound\hyp{}state model was introduced by \cite{Moerdijk1995} as an alternative to coupled channels calculations for the description of Feshbach resonances. In this model the energy of the weakest bound vibrational level of the molecule at zero magnetic field is assumed as a free parameter. The Zeeman shift of this level with respect to the atomic states of the entrance channel is then computed to give an estimate of the location of the Feshbach resonance. The binding energy is then varied to fit the resonant magnetic field to empirical data. For the case of ${}^{6}\textrm{Li}{}^{40}\textrm{K}$ the model was extended to include the hyperfine interaction \cite{Wille2008}, which couples the hyperfine components of the weakest bound vibrational states of the singlet and triplet potentials. This became necessary as the separation between the weakest singlet and triplet bound states is of similar magnitude than the hyperfine coupling. For asymptotic bound states the atomic hyperfine constants are a good approximation. Further, for our case of ${}^{6}\textrm{Li}{}^{40}\textrm{K}$ most Feshbach resonances are narrow. Thus, the coupling between the closed molecular channel and the entrance channel can be neglected. Later, the asymptotic\hyp{}bound\hyp{}state model was extended to include the coupling for ${}^{6}\textrm{Li}{}^{40}\textrm{K}$ and KRb \cite{Tiecke2010,Tiecke2010a} and applied to other cases of Feshbach resonances in heteronuclear mixtures like LiCs \cite{Pires2014}, LiRb \cite{LiZ2008}, and NaK \cite{Park2012}.

For our purposes we resort to the simple model without coupling as described in \cite{Wille2008}. Using the binding energies of the weakest bound states determined by this work, we diagonalize the hyperfine and Zeeman Hamiltonians for varying magnetic field in the molecular basis. To evaluate the case involving the stretched singlet state as discussed in the main text, we choose the total sum of the projection quantum numbers to be $M=-5$. After diagonalization the eigenstate of the resonant molecular level for the Feshbach resonance at $21.56\,\textrm{mT}$ is identified. We then determine the composition of this state in terms of hyperfine components in the uncoupled molecular basis $\ket{S,m_S,m_{I,\mathrm{Li}},m_{I,\mathrm{K}}}$ as shown in Fig.~\ref{fig:ABM-5}. While at zero magnetic field the state has only triplet character, at the location of the resonance the admixture of the singlet state is $52\%$. This represents a favorable condition for a strong transition to excited electronic states of singlet character, in particular the $\textrm{A}^{1}\Sigma^{+}$ state.

\subsection{Mass scaling of spectroscopic data}
To identify the vibrational levels in the $\textrm{A}^{1}\Sigma^{+}$ potential we rely on previous work on polarization labeling spectroscopy in a heat\hyp{}pipe \cite{Grochola2012}. In this work a broad range of rovibrationally excited states was observed and fitted by a Dunham expansion. For our case of addressing deeply bound vibrational states, we can make use of the Dunham coefficients $Y'_{k,l}$ to predict the location of the vibrational lines presented in the main text in Tab.~I. However, since the heat pipe spectroscopy was performed with the ${}^{7}\textrm{Li}{}^{39}\textrm{K}$ isotopologue, the Dunham coefficients need to be mass\hyp{}scaled by the ratio of the reduced masses $\widetilde{\mu}=\mu\left({}^{7}\textrm{Li}{}^{39}\textrm{K}\right)/\mu\left({}^{6}\textrm{Li}{}^{40}\textrm{K}\right)$. We obtain the predicted transition energies from the Dunham expansion with the simple mass scaling \cite{Dunham1932}
\begin{eqnarray*}
E'\left(v',N'\right)&=&h\,	c \sum_{k,l}^{}\widetilde{\mu}^{l+\frac{k}{2}}\,Y'_{k,l}\left(v'+\frac{1}{2}\right)^{k}\left(N'\left(N'+1\right)\right)^{l}\\
 &&-h\,c\,De''\!\!\left(\textrm{X}^{1}\Sigma^{+}\right)-\Delta E''_{\mathrm{ZM}}\,.
\end{eqnarray*}
The Dunham coefficients for the $\textrm{A}^{1}\Sigma^{+}$ potential are determined from transitions originating from the $\textrm{X}^{1}\Sigma^{+}$ potential and referenced to $E''(v''=-1/2)$ \cite{martin2001}. It is then a good approximation to subtract the depth of the electronic ground state potential $De''\!\!\left(\textrm{X}^{1}\Sigma^{+}\right)$, which we take from \cite{Tiemann2009}. Further, a $\Delta E''_{\mathrm{ZM}}=-472.5\,$MHz Zeeman shift of the electronic molecular ground state, which is determined from the above described ABM calculation, is considered to extrapolate the prediction to the Feshbach magnetic field applied during the spectroscopy. The small Zeeman shift of the excited state originating from the rotational excitation is neglected.    

Two uncertainties contribute to the uncertainty of the predicted energies. Firstly, the uncertainty of the depth of the ground state potential is about $6\,$GHz \cite{Tiemann2009}. Secondly, the spectral resolution in \cite{Grochola2012} is given to be $3\,$GHz. Hence, we find that the observed transition lines with a maximal deviation of $10.3\,$GHz from the predicted values are in good agreement with the prediction from mass scaling. Since the level vibrational spacing at $1119\,$nm excitation wavelength is approximately $1.5\,$THz we attain an unambiguous assignment of the vibrational level index. Vibrational lines of the $\textrm{B}^1\Pi$ potential can be excluded, as its minimum is at much higher energies. As discussed in the main text vibrational states from the triplet potentials can be excluded as a resolved hyperfine structure would be expected.

\subsection{Ab\hyp{}initio calculations}
Ab\hyp{}initio calculations of the molecular electronic energy levels do by far not reach an accuracy that is comparable to the typical spectroscopic resolutions, which in our case is about $1\,$MHz. This becomes apparent by using the ab\hyp{}initio potential energy curves (PECs) by \cite{Allouche} for the calculation of the vibrational eigenstates of the $\textrm{A}^{1}\Sigma^{+}$ potential. The transition lines from the ground state calculated from this PECs differ by approximately $6\,$nm from the measured values. Nevertheless, we find it useful to determine trends in molecular properties derived from such calculations as guidelines for experiments, and in the following give an overview.

\subsubsection{Franck\hyp{}Condon factors and transition\hyp{}dipole matrix elements}
\begin{figure}
	\includegraphics[width=8.6cm]{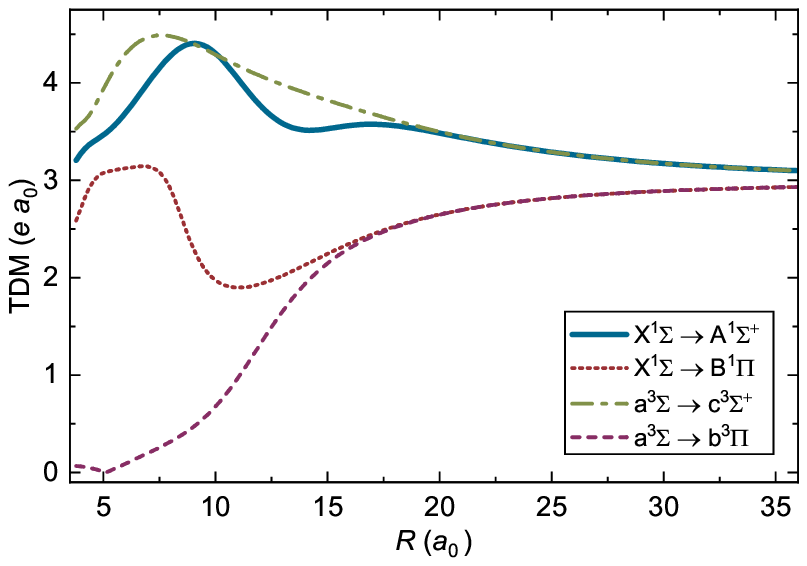}
	\caption{\label{fig:TDMs} Electronic transition dipole matrix elements (TDM) for electronic transitions in the ${}^{6}\textrm{Li}{}^{40}\textrm{K}$ molecule. For the $\textrm{A}^{1}\Sigma^{+}$ the TDM does not strongly vary and slightly increases from the asymptote with decreasing internuclear separation.}
\end{figure}

In order to obtain the transition strengths we calculate the Franck\hyp{}Condon factors for the transitions between the ground state potential as given by \cite{Tiemann2009} and the excited state potentials by \cite{Allouche}. We solve for the vibrational states of these potentials by using the Fourier grid Hamiltonian method \cite{Marston1989}. The results are given in the main text in Fig.~2. As we address low\hyp{}lying vibrational states of the $\textrm{A}^{1}\Sigma^{+}$ potential, it is difficult to discern the Franck\hyp{}Condon points for the transition from the Feshbach state, which extends to an outer turning point at a much larger range, as can be seen in Fig.~1 of the main text. Hence, it is interesting to study the dependence of the electronic  transition\hyp{}dipole matrix element (TDM) with respect to the internuclear separation. This is done by performing an ab\hyp{}initio calculation with the MOLPRO software package \cite{Werner2012,Molpro} using the multi\hyp{}reference configuration interaction (MRCI) method \cite{WK88,KW88,KW92}. The calculation employs Gaussian basis sets with nine valence electrons and relativistic effective core potentials (ECP) for K \cite{LimIS2006} and an accurate all\hyp{}electron basis set for Li \cite{Widmark1990}. The resulting TDMs are shown in Fig.~\ref{fig:TDMs} for the transition to several excited state potentials. For the $\textrm{A}^{1}\Sigma^{+}$ potential the TDM increases from the asymptote towards shorter binding lengths. Hence, transitions to low\hyp{}lying excited vibrational states are expected to slightly increase, which is beneficial to the spectroscopy scheme presented in this work. The use of the $\textrm{A}^{1}\Sigma^{+}$ potential with large TDM is advantageous as compared to transitions involving the $\textrm{b}^{3}\Pi$, where the TDM decreases at short range. This can lead to a significant increase of laser power requirement to drive the excitation to this state.

\subsubsection{Spin\hyp{}orbit coupling}
\begin{figure}
	\includegraphics[width=8.6cm]{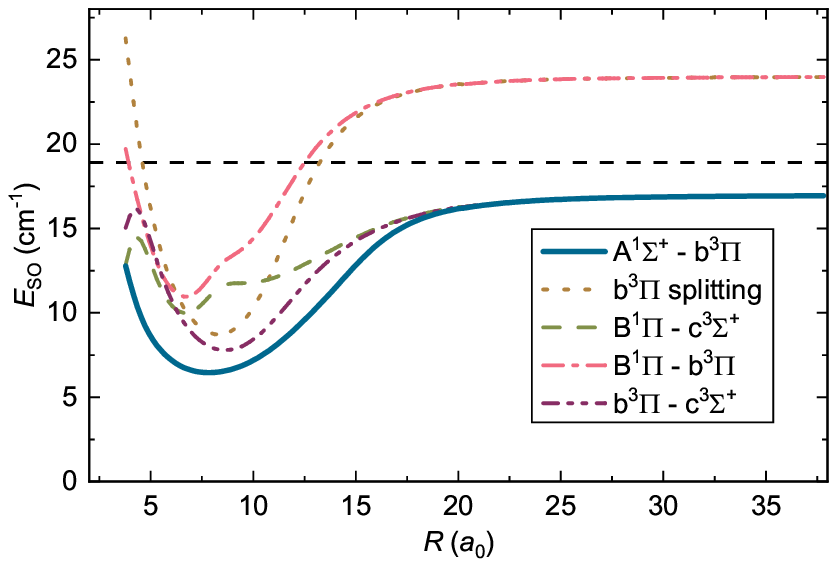}
	\caption{\label{fig:SO-coupling} Spin\hyp{}orbit coupling energies between various electronically excited states of the LiK molecule. The horizontal dashed line indicates the value of the fine structure coupling constant for ${}^{40}\textrm{K}$ atoms \cite{Falke2006}. The discrepancy between the asymptotic value of the spin\hyp{}orbit couplings and the atomic value can be attributed to neglecting the second\hyp{}order spin\hyp{}orbit couplings \cite {Manaa1999,Manaa2002} which are not treated by MOLPRO. 
	}
\end{figure}

Spin\hyp{}orbit coupling between molecular excited states is of great importance for the production of ground state molecules since so far only excited molecular states of mixed singlet and triplet character have been used to address the singlet ground state after starting from a triplet dominated Feshbach state. While in our case of ${}^{6}\textrm{Li}{}^{40}\textrm{K}$ the mixing in the excited state is not desired, mixed states in particular of the $\textrm{A}^{1}\Sigma^{+}$ and $\textrm{b}^{3}\Pi$ potentials are expected to occur, if the rovibrational states of these potentials are nearly degenerate. To illustrate this possibility for low lying vibrational states we calculate the spin\hyp{}orbit coupling constants throughout the full range of the potentials with the MOLPRO software. The calculation makes use of the basis sets previously stated and the ECP approximation. The result in Fig.~\ref{fig:SO-coupling} shows that the spin\hyp{}orbit coupling constant is varying by less than an order of magnitude throughout the range of the potential allowing for the occurrence of mixed states at all binding energies. In general the ab\hyp{}initio PECs do not predict the vibrational states accurately enough to predict the energies at which mixed states occur. For the transition lines to the $\textrm{A}^{1}\Sigma^{+}$ potential reported in our work the mixed character can be excluded, since the expected resolvable hyperfine structure for triplet states is not observed.

\subsubsection{Hyperfine coupling constants}
For none of the observed transitions to the $\textrm{A}^{1}\Sigma^{+}$ potential the hyperfine structure is resolved, even if linear polarized light is used and several hyperfine components are addressed. A larger hyperfine splitting is typically only expected for triplet states, where there is a magnetic dipole\hyp{}dipole interaction between the electronic and nuclear spins or for states of $\Pi$ symmetry, where an internal magnetic field can split the hyperfine components due to the nuclear\hyp{}spin Zeeman effect. To support this argument we investigate the hyperfine coupling strength for the $\textrm{A}^{1}\Sigma^{+}$ state, for which the dominant hyperfine interaction is the nuclear quadrupole interaction. Other contributions to the hyperfine energy of $^1\Sigma$ states, the nuclear spin\hyp{}rotation interaction and the nuclear\hyp{}spin dipolar interaction, typically are much smaller and not considered here. The nuclear quadrupole coupling constant $eQq_{0}(X)$, where $X$ is the atom for which the constant is calculated, can be determined using the linear relationship \cite{Bruna2006}
\begin{eqnarray*}
eQq_{0}(X)[\textrm{MHz}] = 234.9647 \times q_{0}(X)[\textrm{a.u.}] \times Q(X)[\textrm{b}]\ .
\end{eqnarray*}
Here $q_{0}(X)$ is the parallel component of the electric field gradient at nucleus $X$ in atomic units and $Q(X)$ is its nuclear quadrupole moment in barns ($1\,\textrm{b}=10^{-28}\,\textrm{m}$). The nuclear quadrupole moments used to determine the constants are $Q({}^{6}\textrm{Li}) = -8.0\times 10^{-4}\,\textrm{b}$ \cite{Wharton1964} and $Q({}^{40}\textrm{K}) = -75.0\times10^{-3}\,\textrm{b}$ \cite{Teodoro2015}.
The electric field gradients are inferred from ab\hyp{}initio calculations using the DALTON software package with the B3LYP density functional and the 6-311+G* basis sets \cite{Aidas2014,DALTON}. To test the accuracy of this method we calculate the gradient for the $\textrm{X}^{1}\Sigma^{+}$ potential and find the value for ${}^{40}\textrm{K}$ within 5\% of the value in \cite{Aldegunde2017}. At an internuclear distance of $3.95\times 10^{-10}\,$m, which corresponds to the internuclear distance for the deeply bound vibrational states of the $\textrm{A}^{1}\Sigma^{+}$ potential under investigation, we find nuclear quadrupole coupling constants of $eQq_{0}({}^{6}\textrm{Li})=170\,$Hz and $eQq_{0}({}^{40}\textrm{K})=2.97\,$MHz. The small contribution from the lithium nucleus can be neglected. This means that in the strong Feshbach magnetic field in our experiment the hyperfine coupling of the ${}^{6}\textrm{Li}$ nucleus is lifted, and only the coupling for the ${}^{40}\textrm{K}$ nucleus remains dominant. The value for potassium is close to the value for the atomic quadrupole hyperfine interaction of ${}^{40}\textrm{K}$ \cite{Falke2006}. From this value follows a span of $590\,$kHz for the three $\textrm{A}^{1}\Sigma^{+}$ hyperfine states that are addressable with linear polarization in our experiment. This is consistent with the hyperfine structure being not resolvable by the observed linewidth.

%